\documentclass[aps,prc,preprint,groupedaddress,showpacs]{revtex4}
\usepackage{epsfig}
\usepackage{amsmath}

\begin{document}

\preprint{OSU/201-Pcn}

\title{Measurement of the Fusion Probability, P$_{CN}$, for Hot Fusion Reactions}

\author{R. Yanez}
\author{W. Loveland}
\author{J. S. Barrett}
\author{L. Yao}
\affiliation{Department of Chemistry, Oregon State University,
 Corvallis, OR, 97331.}
\author{B.B. Back}
\author{S. Zhu}
\author{T.L. Khoo}
\affiliation{Physics Division, Argonne National Laboratory,
Argonne, IL 60439.}

\date{\today}

\begin{abstract}

{\bf Background:} The cross section for forming a heavy evaporation residue in fusion reactions depends on the capture cross section, the fusion probability, P$_{CN}$, i.e., the probability that the projectile-target system will evolve inside the
fission saddle point to form a completely fused system rather than
re-separating (quasifission), and the survival of the completely fused system against fission.  P$_{CN}$
 is the least known of these quantities.

{\bf Purpose:} To measure P$_{CN}$ for the reaction of 101.2 MeV $^{18}$O ,  147.3 MeV $^{26}$Mg , 170.9 MeV $^{30}$Si  and 195.3 MeV $^{36}$S with  $^{197}$Au. 

{\bf Methods:} We measured the fission fragment angular distributions for these reactions and used the formalism of Back  to deduce the fusion-fission and quasifission cross sections.  From these quantities we deduced P$_{CN}$ for each reaction.

{\bf Results:} The values of P$_{CN}$ for the reaction of 101.2 MeV $^{18}$O ,  147.3 MeV $^{26}$Mg , 170.9 MeV $^{30}$Si  and 195.3 MeV $^{36}$S with  $^{197}$Au are 0.66, 1.00, 0.06, 0.13, respectively.

{\bf Conclusions:} The new measured values of P$_{CN}$ agree roughly with the semi-empirical systematic dependence of P$_{CN}$ upon fissility for excited nuclei.

\end{abstract}

\pacs{25.70.Jj,25.85.-w,25.60.Pj,25.70.-z}

\maketitle

\section{Introduction}
\subsection{Motivation}

The remarkable recent progress in the synthesis of new heavy and superheavy nuclei has been made using fusion reactions.  These reactions can be divided into two prototypical classes, ``cold"  and ``hot" fusion reactions.  In ``cold" fusion reactions, one bombards Pb or Bi target nuclei with heavier projectiles (Ca-Kr) to form completely fused systems with low excitation energies (E*=10-15 MeV), leading to a higher survival (against fission) but with a reduced probability of the fusion reaction taking place due to the larger Coulomb repulsion in the more symmetric reacting system.  (This approach has been used in the synthesis of elements 107-113).  In ``hot" fusion reactions one uses a more asymmetric reaction (typically involving a lighter projectile and an actinide target nucleus) to increase the fusion probability but leading to a highly excited completely fused system (E*=30-60 MeV) with a reduced probability of surviving against fission.  (This approach has been used to synthesize elements 102-118.)

Formally, the cross section for producing a heavy evaporation residue, $\sigma$$_{EVR}$, in a fusion reaction can be written as
\begin{equation}
\sigma _{EVR}=\sum_{J=0}^{J_{\max }}\sigma
_{capture}(E_{c.m.},J)P_{CN}(E*,J) W_{sur}(E*,J)
\end{equation}
where $\sigma _{capture}(E_{c.m.},J)$ is the capture cross section at center of mass energy E$_{c.m.}$ and spin J. P$_{CN}$ is the probability that the projectile-target system will evolve from the contact configuration  inside the
fission saddle point to form a completely fused system rather than
re-separating (quasifission, fast fission). W$_{sur}$ is
the probability that the completely fused system will de-excite by neutron emission rather than fission. For a quantitative understanding of the synthesis of new heavy nuclei, one needs to understand $\sigma _{capture}$, P$_{CN}$, and W$_{sur}$ for the reaction system under study.

The capture cross section is, in the language of coupled channel calculations, the ``barrier crossing" cross section.  It is the sum of the quasifission, fast fission, fusion-fission and fusion-evaporation residue cross sections.  The latter cross section is so small for the systems studied in this work that it is neglected.  In these hot fusion reactions, the capture cross sections have either been measured \cite{bock,clerc,pacheco,prokhorova} or can be predicted, with reasonable accuracy by semi-empirical systematics \cite{sww} or more fundamental calculations  \cite{bain,sarg} .  From a knowledge of $\sigma_{EVR}$ and $\sigma_{capture}$, one can calculate the value of the product W$_{sur}$P$_{CN}$.

The survival probabilities, W$_{sur}$, are calculated using well-established formalisms \cite{vh,zaggyweb} where the principal uncertainty is the values of the fission barrier heights..  Calculations of hot fusion reactions are particularly susceptible to these uncertainties due to the occurrence of multiple chance fission.(The best recent calculations  \cite {adam} of superheavy element fission barrier heights indicate an average discrepancy between data and theory of about 0.4 MeV with the largest discrepancy being about 1.0 MeV.  This latter number roughly translates into an order of magnitude uncertainty in the fission rate).  Nonetheless, the operational procedures for calculating W$_{sur}$ are fairly well understood as well as the dependence of W$_{sur}$ on reaction parameters.

The fusion probability, P$_{CN}$,is the least well-known quantity that determines the evaporation residue cross section \cite{zaggy1}.  Not only is the numerical value of P$_{CN}$ uncertain, but the dependence of P$_{CN}$ on excitation energy \cite{dns,zaggy1,radhika} and the reaction entrance channel is not well established \cite{radhika}.  {\bf It is this quantity, P$_{CN}$, that is the focus of this work.}

\subsection{Reaction Mechanisms}

When a projectile nucleus interacts/reacts with a heavy target nucleus, there are several possible outcomes/mechanisms that come into play.  (Figure 1)  The process of bringing the reacting nuclei into contact and surmounting the interaction barrier is referred to as ``capture" whose probability is reflected in $\sigma$$_{capture}$.  Capture can lead to several different outcomes/mechanisms, i.e., fusion, quasifission, and fast fission.  We briefly summarize the characteristics of each of these dissipative processes as follows: \cite{ct,ngmms}

$\bullet${\bf fusion}- after full momentum transfer, the reacting system evolves inside fission saddle point, resulting in long interaction times and either formation of evaporation residues (fusion-evaporation) (products of complete fusion that de-excite by particle emission) or the formation of mass symmetric fission fragments (fusion-fission)

$\bullet${\bf quasifission}- after full momentum transfer,  and intermediate interaction times, the reacting system {\bf does not} evolve inside the fission saddle point, but re-separates either without significant mass exchange (asymmetric quasifission) or with significant mass transfer (symmetric quasifission).  In any case the fragment angular distributions are more anisotropic than those resulting from fusion-fission.

 $\bullet$ {\bf fast fission}- after full momentum transfer and mass drift, the resulting mono-nucleus fissions because there is no fission barrier, due to the large angular momentum, J, of the system.
 
 The capture cross section, $\sigma$$_{capture}$, is thus
 
 \begin{equation} 
 \sigma_{capture}=\sigma_{fusion}+\sigma_{quasifission} + \sigma_{fast fission}
 \end{equation}

while $\sigma$$_{fusion}$, the fusion cross section, is

 \begin{equation}
 \sigma_{fusion}=\sigma_{fusion-evaporation}+\sigma_{fusion-fission} 
 \end{equation}
 
 where $ \sigma$$_{quasifission}$ is the quasifission cross section and $\sigma$$_{fast fission}$ is the fast fission cross section.  P$_{CN}$ is defined as
 
 \begin{equation}
 P_{CN}=\frac{\sigma _{fusion}}{\sigma _{capture}}=\frac{\sigma
_{capture}-\sigma _{quasifission}-\sigma _{fast fission}}{\sigma _{capture}}
 \end {equation}

\subsection{Quasifission}

The measurement of P$_{CN}$ requires the identification/separation of fusion, quasifission and fast fission (where relevant).  Primarily this task becomes one of identifying quasifission, the re-separation of the contacting nuclei before moving inside the fission saddle point.  There are a series of natural questions about quasifission that are relevant for this problem.  When does quasifission occur?  What are its measurable characteristics?  What are the relevant theoretical models/predictions about quasifission?  What are the experimental data about quasifission?

When does quasifission occur?  Three general, semi-empirical answers to this question are: (a) when the mean fissility of the reacting system, x$_{m}$, exceeds 0.72 \cite{blocki, armbruster} (b) when the product of the atomic numbers of the reacting nuclei, Z$_{1}$Z$_{2}$, exceeds 1600 \cite {wjs, bs, blocki,toke, bock, shen} and (c) when the asymmetry, $\alpha$, of the reacting system is less than the mass asymmetry associated with the Businaro-Gallone point \cite{saxena, pant, berriman, saga2006, ghosh}.  Unfortunately there are known exceptions to each of these general rules, i.e., (a) \cite {berriman, thomas, prasad}  (b) \cite {rafiei, zhang, knya} (c) \cite {knya, appa}.

What are the measurable characteristics of quasifission?  Historically quasifission has been identified by a broadening of the fragment mass distributions caused by the presence of asymmetric mass distributions due to quasifission \cite{ct, ghosh, itkis2010} and by anomalously large fragment anisotropies relative to those expected from fusion-fission \cite{back85,back96}.

Initially one associated symmetric fission with fusion-fission, but it was realized that quasifission could lead to mass symmetric fission fragments \cite{hindeprl, kozulin2009}.  Ultimately one realized the utility of looking at the correlation between fission fragment masses and their angular distributions \cite{hinde2009}.  

What are the relevant theoretical models/predictions about quasifission?  Zagrebaev and Greiner have done a number of calculations of P$_{CN}$ using various approaches \cite{zaggy2,zaggy3}, culminating in some simple semi-empirical predictions for P$_{CN}$(E*) and P$_{CN}$(Z,A) for cold fusion reactions \cite{zaggy1}.  There have been a number of calculations of P$_{CN}$ using the di-nuclear system (DNS) approach for both hot and cold fusion reactions \cite{dns,ngmms,dns3,dns4,dns5} that differ from the Zagrebaev and Greiner predictions \cite{zaggy1}.  A number of calculations of P$_{CN}$ for cold fusion reactions have been made using the ``fusion by diffusion" approach \cite{sww,sww2,sww3,sww4} that differ from both the DNS and Zagrebaev and Greiner approaches.  There have also been a number of attempts \cite{saga2006,sww2,sww4,semi4} to make semi-empirical estimates of P$_{CN}$ using one or another models for $\sigma$$_{capture}$, W$_{sur}$ and using measured values for $\sigma$$_{EVR}$ to get values of P$_{CN}$ for both hot and cold fusion reactions.  Other aspects of quasifission , such as the time scale and the role of deformation effects in the entrance channel have been treated \cite{h1,hinde2009,h3,h4}. Contradictory results have been obtained both experimentally \cite{radhika} and theoretically \cite{zaggy1,dns} as to the expected dependence of P$_{CN}$ upon excitation energy.

What are the experimental data about quasifission/P$_{CN}$?  In Table 1, we attempt to list the current measurements of P$_{CN}$.  The data are sorted by the values of Z$_{1}$Z$_{2}$ which serves as an approximate scaling variable although values of P$_{CN}$ $\leq$ 1 are observed for values of Z$_{1}$Z$_{2}$ $\leq$ 1600.  The data described in Table 1, of necessity, do not include cases where P$_{CN}$ $\ll$ 1 because the quasifission fraction is not measurable, generally, for P$_{CN}$ $\leq$ 0.01.  If one restricts oneself to E* $\sim$40- 50 MeV (to remove the dependence of P$_{CN}$ upon E*), one can discern a rough empirical dependence of P$_{CN}$ upon fissility (Fig. 2).  The data near x$_{eff}$ $\sim$ 0.6 involves $^{48}$Ca projectiles perhaps  reflecting the effects of nuclear structure in the entrance channel upon fusion probability \cite{arm99, hindexx}.  To verify these apparent trends and to allow possible extrapolation/interpolation of these data, 
{\bf it seems as though there is a need for a single measurement of P$_{CN}$ that spans a large range in entrance channel asymmetry at a meaningful excitation energy.}

\subsection {This paper}

In this report, we describe an experimental study that attempts to directly measure P$_{CN}$ in a series of hot fusion reactions and thus to help resolve various issues in predicted values of P$_{CN}$.  Specifically we measured the fission cross section and fragment angular distributions for the reaction of 101.2 MeV $^{18}$O ,  147.3 MeV $^{26}$Mg , 170.9 MeV $^{30}$Si  and 195.3 MeV $^{36}$S with  $^{197}$Au. These systems are described in Table 2 where we show that each system has an excitation energy E* of about 60 MeV. The systems span a range in fissility similar to that covered in the data in Figure 2.    From these data, we have used the method of Back \cite {back85} to deduce the quasifission and complete fusion-fission components of the fragment angular distributions.  We believe (see below) that this method is the best current method of measuring P$_{CN}$.  (We have used this method previously in a study of P$_{CN}$ in cold fusion reactions \cite{radhika}.)  The values of P$_{CN}$ are then compared with current predictions of these quantities.

Some of the systems studied in this work have been studied previously, i.e., the reaction of $^{18}$O with $^{197}$Au \cite{appa, corradi, saga2006}. In \cite{appa}, the fragment angular distributions were measured for the reaction of 78-97 MeV $^{18}$O with $^{197}$Au.  The angular distributions were shown to be consistent with the standard theory of fragment angular distributions, thus indicating that  P$_{CN}$ is 1.  This system can thus be a check on the reproducibility of the experimental measurements and their interpretation. Corradi, et al., \cite {corradi} measured the yields of the Fr evaporation residues from the $^{18}$O + $^{197}$Au reaction for beam energies of 75-130 MeV.  Sagaidak et al. \cite{saga2006} took these data on evaporation residue production cross sections and compared them to the predictions of the computer code HIVAP assuming P$_{CN}$ was 1.  For a best fit to the data, they had to assume a reduction in the fission barrier height of 15$\%$, which could also, as the authors point out, be taken as a need to decrease P$_{CN}$.   Another related study \cite{back95a} was that of the fragment angular distributions in the reaction of 185 MeV $^{32}$S with $^{197}$Au  where larger fragment anisotropies were observed than predicted by a rotating liquid model of the fissioning nucleus and the excitation energy E* was 60 MeV.The fragment angular distributions were measured radiochemically for the interaction of $^{11}$B, $^{12}$C, $^{14}$N, and $^{16}$O with $^{197}$Au \cite{vts}.  The data are well described by the standard theory of fission fragment angular distributions \cite{vh}, except that there were some difficulties due to the occurrence of incomplete fusion at higher bombarding energies and the extracted values of the mean spin of the fissioning systems were low for reactions near the Coulomb barrier.  

In summary, previous work supports the idea that P$_{CN}$ is 1 for the most asymmetric systems involving the interaction of lighter projectiles with $^{197}$Au.

\section {Experimental Methods}

The experiment was carried out in the ATSCAT scattering chamber at the ATLAS accelerator facility at the Argonne National Laboratory.  The experimental setup is shown in Figure 3.  Beams of $^{18}$O, $^{26}$Mg, $^{30}$Si, and $^{36}$S struck a  $^{197}$Au target  mounted at the center of the scattering chamber.  We assumed that all ion charges equilibrated in the 0.25 mg/cm$^{2}$ $^{197}$Au target and the equilibrium charge values \cite{shima} were used in calculating beam doses.  The beam intensity was monitored in two ways : (a) by a deep Faraday cup at the periphery of the chamber and (b) a pair of silicon monitor detectors (r = 2.00 mm) mounted at 15$^{\circ}$ with respect to the beam axis at a distance of 412.75 mm from the target.  A voltage of +9 kV was applied to the target to suppress the emission of energetic $\delta$ electrons. The beam intensities ranged from 2 to 3 x 10$^{10}$ p/s.  All beam energies used herein are the center of target beam energies calculated using SRIM \cite{srim}.

On one side of the beam, we mounted an array of Si detectors whose positions were fixed during all measurements. The angles of these detectors were 73$^{\circ}$, 78$^{\circ}$, 83$^{\circ}$ and 89$^{\circ}$.  On the other side of the beam there were two independently movable arrays, one at forward angles and one at backward angles,  The forward array consisted of three detectors nominally separated in angle by 5$^{\circ}$. The backward array consisted of seven detectors nominally separated by 5$^{\circ}$ from each other.  All the array detectors had an area of 300 mm$^{2}$ and were positioned at $\sim$ 320 mm from the target.  For each projectile-target combination, six positions of the forward/backward arrays were used.  For the backward array angles of 82 - 172 $^{\circ}$ were sampled while the forward array was used for measurements at angles of 53 - 75 $^{\circ}$.

Time information was measured for each Si detector relative to the Linac pulse structure.  The beam was bunched into packets with a FWHM of 0.71 ns. The average time between beam bursts was 82 ns. From the particle time of flight and energy, the mass of the fragment was calculated.

Energy calibrations of each Si detector were performed using the response of the detectors to $^{252}$Cf fission fragments  and elastically scattered beam from a $^{197}$Au target.  A correction for pulse height defect was applied to fission fragments striking the detectors, using the response of the detectors to $^{252}$Cf fission fragments. \cite{skw} 

The reaction of heavy ions with $^{197}$Au can lead to elastic scattering, inelastic scattering, deep inelastic scattering, fusion-fission, quasifission and fusion-evaporation residue formation.  (The cross section for evaporation residue formation is small in comparison with the other processes and will be neglected in this discussion).  Fusion-fission and quasifission events were isolated from the other types of events by analyzing the E vs A response of each detector.

\section{Results}
\subsection{Capture Cross Sections}

The singles fission data at backward angles was integrated using data with $\theta_{lab}$ $\leq$172$^{\circ}$.  The total cross section was deduced from these differential cross sections  by the simple assumption that the fission fragments were emitted in a plane perpendicular to the total angular momentum vector, i.e., the fragment angular distribution is given by 
\begin{equation}
W(\theta) = (2\pi^{2}sin\theta)^{-1}
\end{equation}

The resulting capture-fission cross sections are shown in Fig. 4 and Table II.  In Fig. 4, we also show a previous measurement of $\sigma_{capture-fission}$ for the reaction of 97 MeV $^{18}$O + $^{197}$Au \cite{appa}.  We also show a measurement of $\sigma_{capture-fission}$ for the reaction of 185 MeV $^{32}$S + $^{197}$Au \cite{bigback} where the measured cross section of 250 $\pm$ 15 mb has been scaled (multiplied by) the ratio of the Bass model fusion cross sections (1.55) for the 185 MeV $^{32}$S + $^{197}$Au and the 195.3 MeV $^{36}$S + $^{197}$Au cross sections.  We also show the predictions for these cross sections obtained using two statistical model codes for heavy element reactions, HIVAP \cite{hivap} and the coupled channels approach of Zagrebaev \cite{zaggyweb}.  The discordance amongst the measurements and the predictions is discouraging although this situation is consistent with previous evaluations of factors of 2 - 10 discrepancies in calculating $\sigma$$_{capture}$ \cite{wdltex, zaggyoga}.

\subsection{Fragment Angular Distributions}
The fission fragment angular distributions were measured using the individual Si detectors and are shown in Figure 5.   

It has been shown \cite{back85,keller} that for some reactions that a
significant fraction of the fission events result from ``quasifission'' as
well as ``true complete fusion''. Quasifission is the process where the
interacting nuclei merge to form a mononucleus but the system does not
evolve inside the fission saddle point.  For the
purpose of estimating heavy element production by complete fusion, one must
separate the contributions of quasifission and true complete fusion in the
data. Using the methods outlined in ref \cite{back85,keller} which depend on
analyzing the shape of the fission fragment angular distributions, we have
attempted to estimate the relative contributions of quasifission and
complete fusion to the observed cross sections. 

The authors of \cite{back85,keller}
studied the angular distributions for a large number of reactions. They concluded that
one could decompose the observed fission angular distributions into two
components, one due to complete fusion and the other due to quasifission. \
The complete fusion component has an angular distribution characterized by
values of the effective moment of inertia, $\Im _{eff}$, as taken from the rotating liquid drop model \cite{cps,sierk} for J $\leq$ J$_{CN}$ 
while the quasifission component has

\begin{equation}
\frac{\Im _{0}}{\Im _{eff}}=\text{1.5 \ }J > J _{cn}
\end{equation}

Figure 12 in \cite{betty} shows the shapes associated with various values of $\Im _{0}$/$\Im _{eff}$ .  

In these equations, $\Im _{0}$ is the moment of inertia of a spherical nucleus
with the same A value, complete fusion is assumed to occur for spins
0$\leq J \leq J _{cn}$ and quasifission is assumed to occur for spins J $ 
\mbox{$>$}%
J _{cn}$. We fitted the observed fission fragment angular distributions allowing the maximum angular momentum associated with complete fusion, J$_{CN}$,  to be a free parameter determined in the calculation. \ (J$_{\max }$ was determined from the sum of the complete fusion-fission and quasifission cross sections using a sharp cutoff approximation.) \ We used the familiar expressions for the fission fragment angular distributions\cite{hbm},

\begin{equation} 
\begin{split}
W(\theta )=&\sum\limits_{J=0}^{J_{CN}}\frac{(2J+1)^{2}exp
[-(J+1/2)^{2}sin ^{2}\theta /4K_{0}^{2}(FF)]J_{0}[i(J+1/2)^{2}sin ^{2}\theta
/4K_{0}^{2}(FF)]}{erf[(J+1/2)/(2K_{0}^{2}(FF))^{1/2}]}\\
&+\sum\limits_{J=J_{CN}}^{J_{\max }}\frac{(2J+1)^{2}exp [-(J+1/2)^{2}sin
^{2}\theta /4K_{0}^{2}(QF)]J_{0}[i(J+1/2)^{2}sin ^{2}\theta /4K_{0}^{2}(QF)]}{{erf}[(J+1/2)/(2K_{0}^{2}(QF))^{1/2}]}\\
\end{split}
\end{equation}
neglecting the spins of the target and projectile, where J$_{0}$ is the zero order Bessel function with imaginary
argument and the error function erf[(J+1/2)/(2K$_{0}^{2})^{1/2}]$ is defined
as 
\begin{equation}
erf(x) =(2/\pi ^{1/2})\int\limits_{0}^{x}\exp (-t^{2})dt
\end{equation}

The parameter K$_{0}^{2}$ is defined as 
\begin{equation}
K_{0}^{2}=T\Im _{eff}/\hbar ^{2}
\end{equation}
\begin{equation}
\frac{1}{\Im _{eff}}=\frac{1}{\Im _{\parallel }}-\frac{1}{\Im _{\perp }}
\end{equation}
where the nuclear temperature at the saddle point T is given as 
\begin{equation}
T=\left[ \frac{E^{\ast }-B_{f}-E_{rot}-E_{\nu}}{A/8.5}\right] ^{1/2}
\end{equation}
and $\Im _{\parallel }$ and $\Im _{\perp }$ are the moments of inertia for rotations around the axis parallel and perpendicular to the nuclear symmetry axis, respectively. B$_{f}$ , E$_{rot}$ and E$_{\nu}$ are the fission barrier, the rotational energy of the system and the energy lost in the emission of pre-fission neutrons.   This later quantity is taken from estimates from \cite{zaggyweb}.
The assumption that 
\begin{equation}
\frac{\Im _{0}}{\Im _{eff}}=\text{1.5 \ } for J \succ J _{cn}
\end{equation}
for quasifission is arbitrary.  The value of this ratio is greater than that observed in any complete fusion-fission reaction for this fissility \cite{back85} but the actual value of this ratio is not well established.  

In fitting the angular distribution data, one uses the measured value of $\sigma_{capture}$, and $K_0^2$ values calculated from equation 9 and varies J$_{CN}$  until a minimum in the reduced chi-square,$\chi^{2}_{\nu}$, is achieved.  Then
\begin{equation}
\frac{J_{CN}^{2}}{J_{\max }^{2}}=\frac{\sigma _{CN}}{\sigma
_{capture}}=P_{CN}
\end{equation}

The final fits to the measured angular distributions are shown in Figure 5.  The deduced values of P$_{CN}$ for the reaction of 101.2 MeV $^{18}$O ,  147.3 MeV $^{26}$Mg , 170.9 MeV $^{30}$Si  and 195.3 MeV $^{36}$S with  $^{197}$Au are 0.66, 1.00, 0.06, 0.13, respectively. For all cases, the $\chi^{2}_{\nu}$ values were statistically significant at the 95 $\%$ level.\cite{lyon}  It is difficult to make meaningful estimates of the uncertainties in the deduced values of P$_{CN}$ given the fundamental systematic uncertainties in $\frac{\Im _{0}}{\Im _{eff}}$ and thus in $K_0^2$.

\section{Discussion}

From analyzing our data for the reaction of 101.2 MeV $^{18}$O with $^{197}$Au, we deduced a value of P$_{CN}$ of 0.66.  The authors of \cite{appa} found their angular distributions for the reaction of 97 MeV $^{18}$O with $^{197}$Au were consistent with the standard theory of angular distributions, presumably indicating P$_{CN}$ = 1.  Sagaidak et al. \cite{saga2006} analyzed the evaporation residue data of Corradi et al. \cite{corradi} for 75-130 MeV $^{18}$O + $^{197}$Au and found the fission barriers had be lowered by a factor of 0.75 to fit the data.  They noted a similar situation in the $^{19}$F + $^{197}$Au reaction where the similar results (k$_{f}$ = 0.85) could also be accounted for if P$_{CN}$ =0.75.  Another relevant observation is that of Viola, Thomas and Seaborg \cite{vts} who studied the fragment angular distributions in the closely related $^{16}$O + $^{197}$Au reaction and who found they were unable to describe the distributions with standard methods.  Given all this information, our measured value of P$_{CN}$ = 0.66 for the reaction of 101.2 MeV $^{18}$O with $^{197}$Au seems reasonable.

Back et al \cite{bigback} have previously measured and analyzed the fission fragment angular distributions for the reaction of 185 - 225 MeV $^{32}$S + $^{197}$Au.  They concluded that the value of $\frac{\Im _{0}}{\Im _{eff}}$ needed to describe the data was significantly different (1.5 - 2 x) from that predicted by the rotating liquid drop model .  That is qualitatively consistent with our finding of P$_{CN}$ = 0.13 for the $^{36}$S + $^{197}$Au system.

In Figure 6, we compare our measured values of P$_{CN}$ (from this work) to the systematics of P$_{CN}$ values \cite{wdltex} for systems with E*= 40 - 50 MeV.  (Strictly speaking , since all the systems studied in this work involve E* $\sim$ 60 MeV, we should scale the measured values to E*= 40 - 50 MeV.  However, because of the controversy \cite{zaggy1,dns} about how to do this scaling, we are just plotting our unscaled data.)

All the new data values of P$_{CN}$ are within an order of magnitude of the systematic trend of the previous data.  (That is consistent with general predictions of the uncertainties in our knowledge of P$_{CN}$ \cite{zaggyoga}. ) Zagrebaev and Greiner have suggested \cite{zaggy1} that for data where E* $\le$ 40 MeV, that P$_{CN}$ (at constant E*) might show a simple behavior 
\begin{equation}
P_{CN}=\frac{1}{1+\exp \left[ \frac{Z_{1}Z_{2}-\zeta }{\tau }\right] }
\end{equation}
where $\zeta$ = 1760 and $\tau$ = 45.  This expression is intended only to represent P$_{CN}$ for cold fusion reactions with $^{208}$Pb or $^{209}$Bi.  As seen in Figure 7, this expression overestimates the values of P$_{CN}$ for hot fusion reactions.

The ``fusion by diffusion" model \cite{sww} includes a formalism for calculating P$_{CN}$ that should be applicable to hot fusion reactions.  The predictions of that formalism are also shown in Figure 7.  This model gives estimates of P$_{CN}$ that are lower than the measured values but which generally describe the dependence of P$_{CN}$ upon fissility.

Also included in Figure 7 is a simple empirical representation of the data (as a dotted line, i.e.,for 
 x$_{eff}$ $\le$ 0.58,  P$_{CN}$ =1.   For x$_{eff}$ $\geq$  0.58,P$_{CN}$=exp (-26.8(x$_{eff}$-0.58)).  
 
 Siwek-Wilczynska et al. \cite{sww4} have proposed a parameterization of P$_{CN}$ in the form of an equation
 \begin{equation}
 log_{10}(P_{CN})= -(z/a)^{k}
 \end{equation}
 where a $\sim$ 145 and k = 3.0.  The variable z is given as
 \begin{equation}
 z=\frac{Z_{1}Z_{2}}{\left ( A_{1}^{1/3} +A_{2}^{1/3} \right )}
 \end{equation}
While this function is intended to describe situations where E* is about 10 MeV above the barrier, it appears do a respectable job of representing the P$_{CN}$ values.

If we look carefully at the data with x$_{eff}$ $\sim$ 0.6, we see a variation of an order of magnitude in P$_{CN}$ with an approximately constant value of x$_{eff}$.  It seems clear that x$_{eff}$ is not adequate as a single scaling variable to determine P$_{CN}$, as we had thought previously. \cite{arm99}.

\section{Conclusion}

What have we learned in this study?  We have measured values of P$_{CN}$ for four new reactions.   The values of P$_{CN}$ for the reaction of 101.2 MeV $^{18}$O ,  147.3 MeV $^{26}$Mg , 170.9 MeV $^{30}$Si  and 195.3 MeV $^{36}$S with  $^{197}$Au are 0.66, 1.00, 0.06, 0.13, respectively.  These reactions span a range of fissility used previously to compile a data set of P$_{CN}$ values for hot fusion reactions.  The new data are in rough agreement with previous measurements supporting the general dependence of P$_{CN}$ upon fissility. Some current models for estimating P$_{CN}$ are not adequate for quantitatively specifying P$_{CN}$.  The effective fissility, x$_{eff}$, is a rough scaling variable for P$_{CN}$ but systems with similar x$_{eff}$ can have P$_{CN}$ values differing by as much as an order of magnitude.

\begin{acknowledgments}

This work was supported in part by the
Director, Office of Energy Research, Division of Nuclear 
Physics of the Office of High Energy and Nuclear Physics 
of the U.S. Department of Energy
under Grant DE-FG06-97ER41026 and contract No. DE-AC02-06CH11357.
\end{acknowledgments}

\newpage
\begin{table}[ht]
\begin{center}
\vspace{0.5cm}
\caption{Measurements of P$_{CN}$.  The methods are angular distribution measurements (AD), mass distribution measurements (MY), and mass-angle measurements (MAD).}
\begin{ruledtabular}
\begin{tabular}{cccccccccccc}
Proj.&Target&CN&E$_{c.m.}$(MeV) &E*(MeV)&Z$_{1}$Z$_{2}$&$\alpha$&$\alpha$$_{BG}$&x$_{eff}$&P$_{CN}$& Ref&Method   \\
\hline
$^{11}$B&$^{204}$Pb&$^{215}$At&48-60&31-43&410&0.898&0.761&0.325&1-1&\cite{appa}&AD \\
$^{16}$O&$^{186}$W&$^{202}$Pb&70-121&48-100&592&0.842&0.765&0.42&1-1&\cite{knya}&MAD \\
$^{18}$O&$^{197}$Au&$^{215}$At&71-89&39-56&632&0.833&0.788&0.413&1-1&\cite{appa}&AD \\
$^{19}$F&$^{208}$Pb&$^{227}$Pa&101-174&51-124&738&0.833&0.816&0.459&0.78-0.83&\cite{back85}&AD \\
$^{24}$Mg&$^{208}$Pb&$^{232}$Pu&126-188&52-114&984&0.793&0.847&0.549&0.64-0.71&\cite{back85}&AD \\
$^{48}$Ca&$^{144}$Sm&$^{192}$W&141-167&38-64&1080&0.5&0.756&0.544&1-1&\cite{knya}&MAD \\
$^{28}$Si&$^{208}$Pb&$^{236}$Cm&141-229&50-138&1148&0.763&0.862&0.597&0.37-0.63&\cite{back85}&AD \\
$^{26}$Mg&$^{248}$Cm&$^{274}$Hs&119-146&37-64&1152&0.81&0.886&0.572&0.6&\cite{imit}&MY \\
$^{32}$S&$^{182}$W&$^{214}$Th&141-221&56-136&1184&0.701&0.851&0.613&0.14-0.51&\cite{keller}&AD \\
$^{48}$Ca&$^{154}$Sm&$^{202}$Pb&139-185&49-95&1240&0.525&0.813&0.594&0.55-0.94&\cite{knya}&MAD \\
$^{40}$Ca&$^{154}$Sm&$^{194}$Pb&139-158&56-75&1240&0.588&0.828&0.633&0.89-0.98&\cite{knya}&MAD \\
$^{32}$S&$^{208}$Pb&$^{240}$Cf&172-217&66-111&1312&0.733&0.875&0.641&0.45-0.46&\cite{back85}&AD \\
$^{36}$S&$^{238}$U&$^{274}$Hs&153-173&36-56&1472&0.737&0.896&0.647&0.043-0.3&\cite{imit}&MY \\
$^{50}$Ti&$^{208}$Pb&$^{258}$Rf&184-202&14-33&1804&0.612&0.899&0.725&0.02-0.19&\cite{radhika}&AD \\
$^{48}$Ca&$^{238}$U&$^{286}$Cn&185-215&26-56&1840&0.664&0.911&0.713&0.00025-0.125&\cite{koz2010}&MY \\
$^{64}$Ni&$^{238}$U&$^{302}$120&267-300&30-63&2576&0.576&0.939&0.867&0.021-0.047&\cite{koz2010}&MY \\
\label{table1}
\end{tabular}
\end{ruledtabular}
\end{center}
\end{table}

\newpage
\begin{table}[ht]
\begin{center}
\vspace{0.5cm}
\caption{Characteristics of Reactions Studied in this work}
\begin{ruledtabular}
\begin{tabular}{ccccccccccc}
Proj.&Target&CN&E$_{c.m.}$(MeV) &E*(MeV)&Z$_{1}$Z$_{2}$&$\alpha$&$\alpha$$_{BG}$&x$_{eff}$& $\sigma_{capture-fission}$ (mb)   P$_{CN}$ \\
\hline
$^{18}$O&$^{197}$Au&$^{215}$At&92.8&60.5&632&0.833&0.788&0.413&834 $\pm$ 4&0.66 \\
$^{26}$Mg&$^{197}$Au&$^{223}$Pa&130.1&60.4&948&0.767&0.833&0.524& 749 $\pm$ 6 &1.0 \\
$^{30}$Si&$^{197}$Au&$^{227}$Np&148.3&60.2&1106&0.736&0.849&0.572& 770 $\pm$ 10 &0.06 \\
$^{36}$S&$^{197}$Au&$^{233}$Am&165.1&60.1&1264&0.691&0.860&0.604& 748 $\pm$ 10 &0.13 \\
\label{table1}
\end{tabular}
\end{ruledtabular}
\end{center}
\end{table}

\newpage 
\begin{figure}[tbp]
\begin{minipage}{30pc}
\begin{center}
\includegraphics [width=150mm]{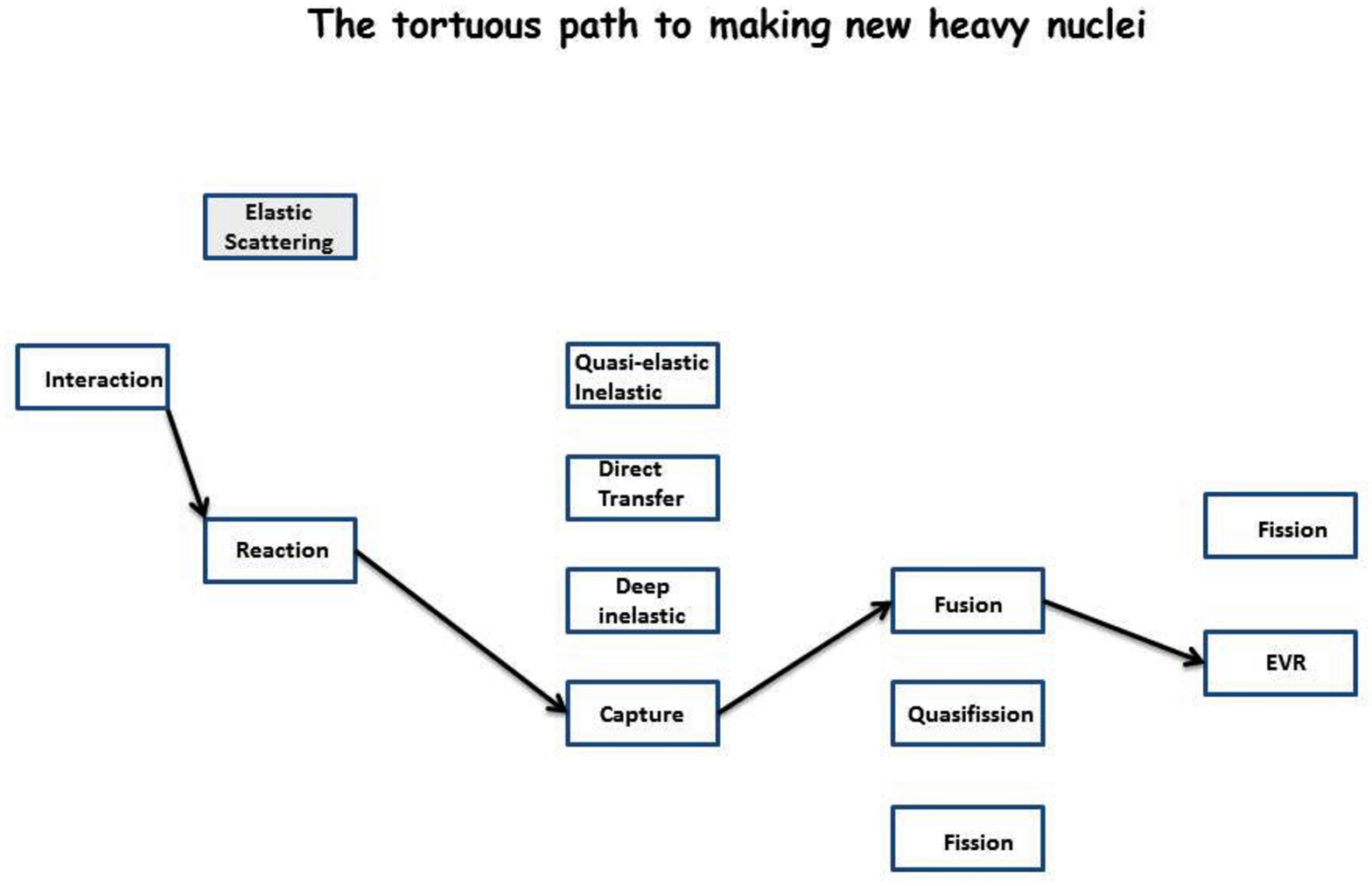}
\end{center}
\caption{Schematic diagram of the path to synthesize new heavy nuclei, showing the reaction mechanisms involved. }
\label{fig1}
\end{minipage}
\end{figure}

\newpage 
\begin{figure}[tbp]
\includegraphics [scale=0.70]{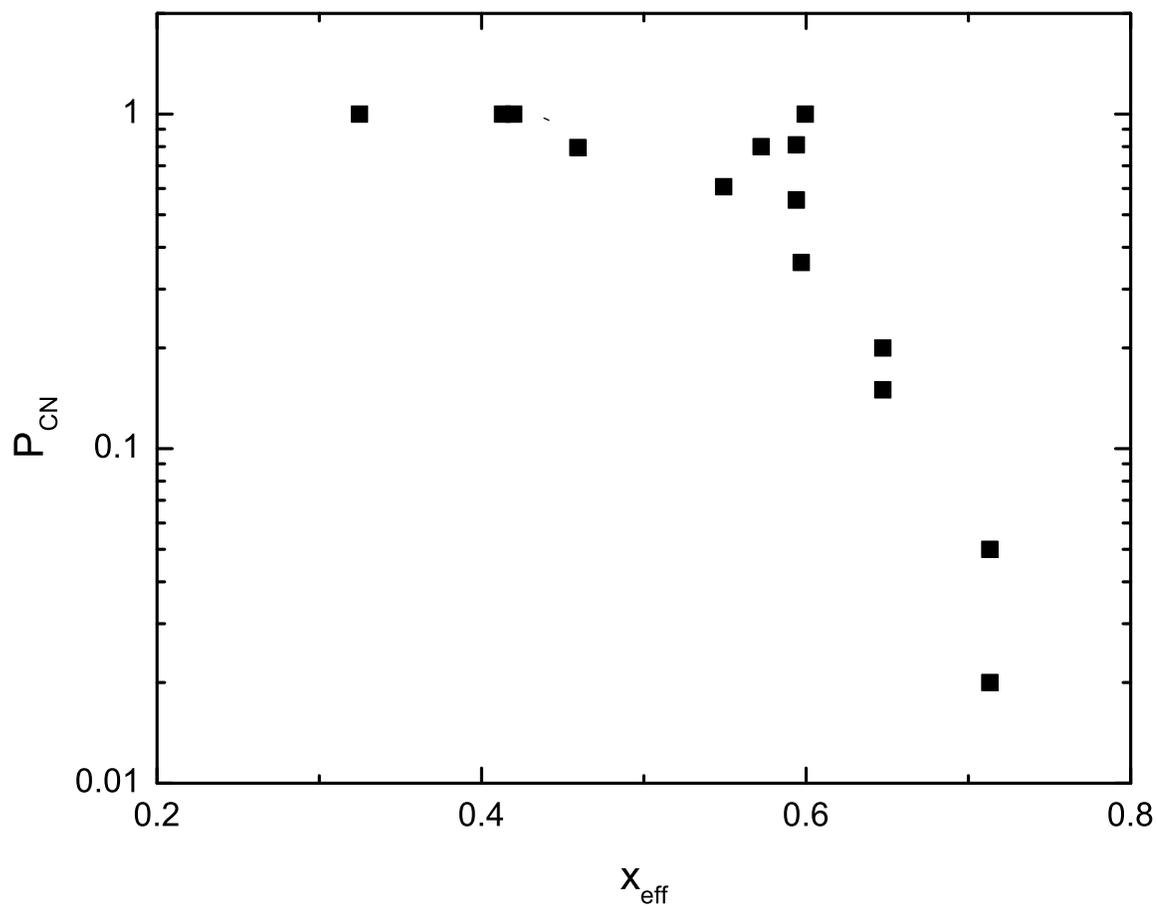}
\caption{A plot of measured values of P$_{CN}$ vs. the scaling parameter, the effective fissility, x$_{eff}$.  The data are from Table I for systems where E* $\sim$ 40-50 MeV. }
\label{fig2}
\end{figure}

\newpage 
\begin{figure}[tbp]
\begin{minipage}{30pc}
\begin{center}
\includegraphics [width=100mm]{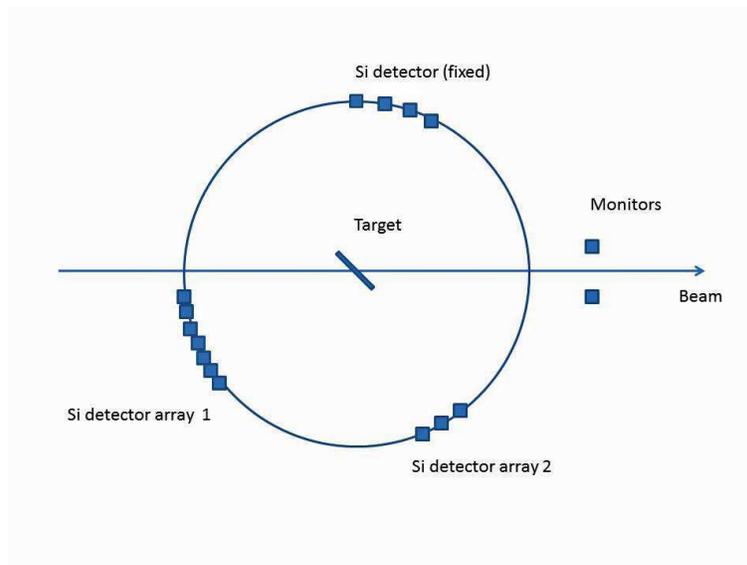}
\end{center}
\caption{(Color online)Schematic diagram of the experimental apparatus. }
\label{fig3}
\end{minipage}
\end{figure}

\newpage 
\begin{figure}[tbp]
\begin{minipage}{30pc}
\begin{center}
\includegraphics [width=120mm]{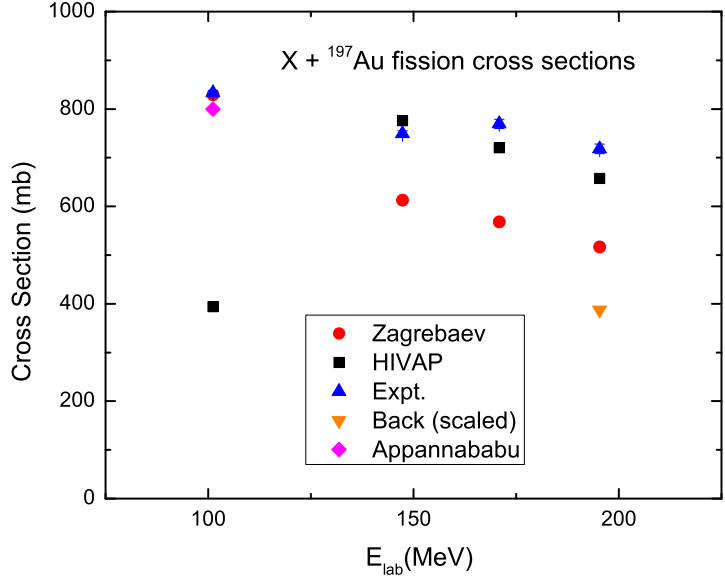}
\end{center}
\caption{ (Color online) Measured capture-fission cross sections for the reactions studied in this work along with previous measurements \cite{appa,bigback} and statistical model estimates of the these cross sections. \cite{hivap, zaggyweb}  The lab energies for the $^{18}$O, $^{26}$Mg, $^{30}$Si and $^{36}$S reactions were 101.2, 147.3, 170.9, and 195.3 MeV, respectively.}
\label{fig4}
\end{minipage}
\end{figure}

\newpage 
\begin{figure}[tbp]
\begin{minipage}{40pc}
\begin{center}
\includegraphics [width=190mm]{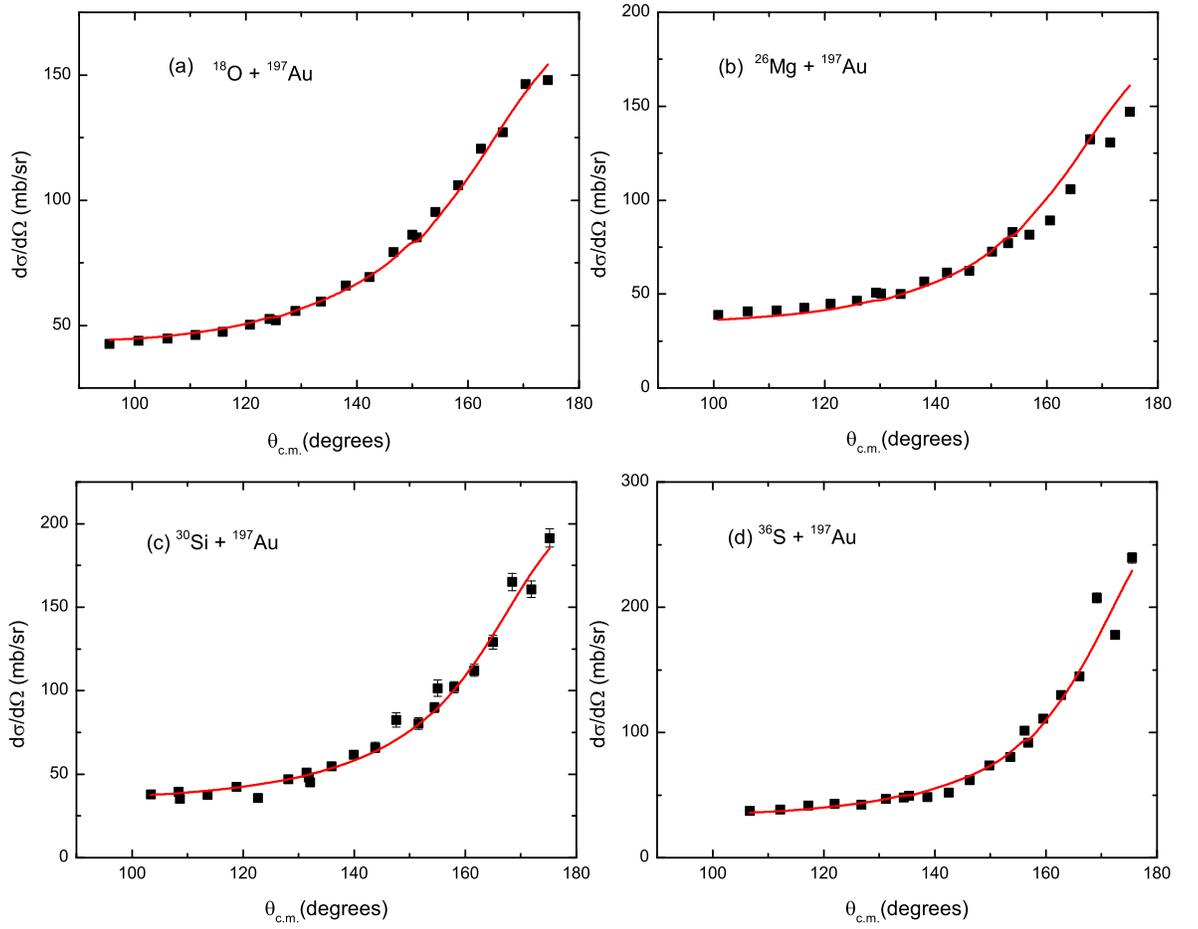}
\end{center}
\caption{ (Color online) Measured fission fragment angular distributions for the reactions studied in this work and the resulting fits to the distributions to resolve complete fusion-fission from quasifission.}
\label{fig5}
\end{minipage}
\end{figure}

\newpage 
\begin{figure}[tbp]
\begin{minipage}{40pc}
\begin{center}
\includegraphics [width=150mm]{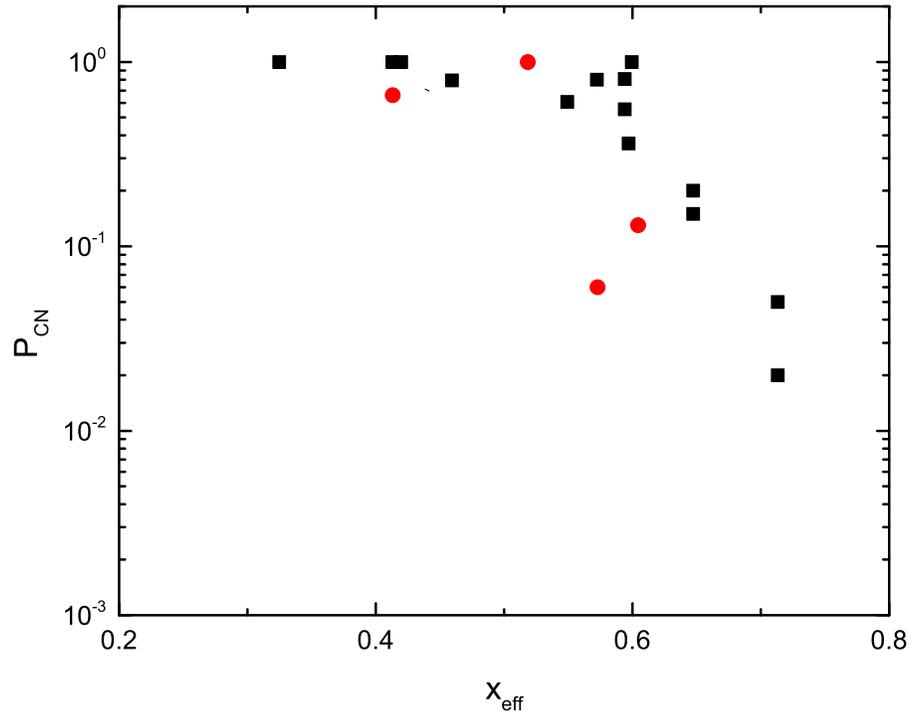}
\end{center}
\caption{ (Color online) Comparison of the measurements from this work with the systematic dependence of P$_{CN}$ upon fissility.  The red circles are the data from this work while the black squares represent previous measurements.}
\label{fig5}
\end{minipage}
\end{figure}

\newpage 
\begin{figure}[tbp]
\begin{minipage}{30pc}
\begin{center}
\includegraphics [width=120mm]{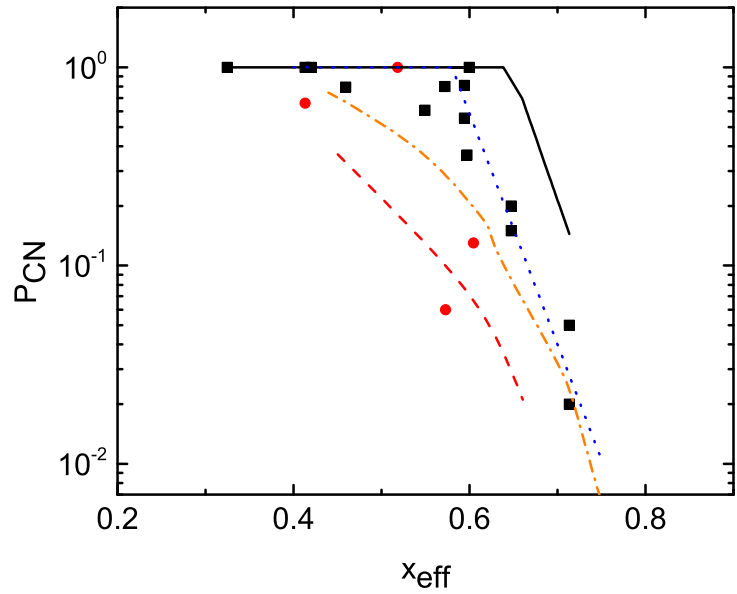}
\end{center}
\caption{ (Color online) Comparison of the measurements of P$_{CN}$ with four models for P$_{CN}$.  Solid line \cite{zaggy1}, dashed line \cite{sww}, dot-dash line is from \cite{sww4}, dotted line is simple fit to data.}
\label{fig5}
\end{minipage}
\end{figure}

\end{document}